\documentclass{PoS}

\title{First ADS analysis of
$B^-\to D^0K^-$ decays\\
in hadron collisions}

\ShortTitle{ADS analysis of $B \to DK$}

\author{\speaker{Paola Garosi}\\
        University of Siena and INFN Pisa\\
        E-mail: \email{paola.garosi@pi.infn.it}}

\abstract{Measurements of branching fractions and $CP$-asymmetries of $B^- \to D^0 
K^-$ modes allow a  theoretically-clean extraction of the CKM angle $\gamma$.
The method proposed by Atwood, Dunietz and Soni (ADS) makes use of a decay chain where color and Cabibbo suppression interfere, which produces large $CP$-violating asymmetries.
The CDF experiment reports the first measurement at a hadron collider of branching fractions and $CP$-asymmetries of suppressed $B^- \to D^0 h^-$ signals, where $h$ is $\pi$ or $K$. Using 5.0 fb$^{-1}$ of data we found a combined significance exceeding 5$\sigma$ and we determined the ADS parameters with accuracy comparable with $B$-factories.}

\FullConference{The 13th International Conference on B-Physics at Hadron Machines - Beauty2011,\\
		April 04-08, 2011\\
		Amsterdam, The Netherlands}

\begin{document}
\section{Introduction}
The measurement of the CKM matrix elements plays a central role both to test the Standard Model consistency and to probe New Physics scenarios. 
The complex phase of the CKM matrix leads to $CP$ violation in weak processes. Observables are written in terms of the angles $\alpha$, $\beta$ and $\gamma$ of the ``Unitarity Triangle", obtained from the unitarity condition of the CKM matrix~\cite{ref:CKM}.
While the resolutions on $\alpha$ and $\beta$ reached a good level of precision, the measurement of $\gamma= arg (- V_{ud} V^*_{ub} / V_{cd} V^*_{cb})$ is still limited by the smallness of the branching ratios involved in the processes and its uncertainty varies between $11$ and  $25$ degrees, depending on the method used to combine the experimental results~\cite{ref:hfag,ref:utfit,ref:ckmfitter}.

Among the various methods for the $\gamma$ measurement, those which make use of the tree-level dominated $B^- \to D K^-$ decays (where $D$ labels either $D^0$ or $\overline{D}^0$ mesons) have the smallest theoretical uncertanties~\cite{ref:glw1,ref:ads1,ref:ggsz}. In fact $\gamma$ appears as the relative weak phase between two amplitudes, the favored $b \to c \bar{u} s$ of the $B^- \to D^0 K^-$ (whose amplitude is proportional to $V_{cb} V_{us}$) and the color-suppressed $b \to u \bar{c} s$ of the $B^- \to \overline{D}^0 K^-$ (whose amplitude is proportional to $V_{ub} V_{cs}$). The interference between $D^0$ and $\overline{D}^0$ decaying into the same final state leads to a measurable $CP$-violating effect.

According to the final state of the $D$, we have the following methods
\textit{GLW (Gronau-London-Wyler) method}~\cite{ref:glw1,ref:glw2}, which uses $CP$ eigenstates of $D^0$, as $D^0_{CP^+} \rightarrow K^+ K^-, \pi^+ \pi^-$ and $D^0_{CP-} \rightarrow K^0_s \pi^0, K^0_s \phi, K^0_s \omega$;
\textit{ADS (Atwood-Dunietz-Soni) method}~\cite{ref:ads1,ref:ads2}, which uses the doubly Cabibbo suppressed mode $D^0 \rightarrow K^+ \pi^-$ and
\textit{GGSZ (or Dalitz) method}~\cite{ref:ggsz,ref:ads2}, which uses three body decays of $D^0$, as 
$D^0 \rightarrow~K^0_s \pi^+ \pi^-$.

All mentioned methods require no tagging or time-dependent 
measurements, and many of them only involve charged particles in 
the final state. They are therefore particularly well-suited to analysis in a hadron 
collider environment, where the large production of $B$ mesons can be exploited.
The use of a specialized trigger based on online detection of a secondary vertex (SVT trigger~\cite{ref:trigger}) allows the selection of pure $B$ meson samples.

We will describe in more details the ADS and GLW methods, for which CDF reports the first results in hadron collisions.

%

\section{The Atwood-Dunietz-Soni method}
In the ADS method~\cite{ref:ads1,ref:ads2} the interference between these two decay channels is studied:
$B^- \to D^0 K^-$ (\textit{color favored}), with $D^0 \to K^+ \pi^-$ (\textit{doubly Cabibbo suppressed})
and $B^- \to \overline{D}^0 K^-$ (\textit{color suppressed}), with $ \overline{D}^0 \to K^+ \pi^-$ (\textit{Cabibbo favored}).
Since $D^0$ and $\overline{D}^0$ are indistinguishable, the final state $[K^+ \pi^-]_D K^-$ is reconstructed and the direct $CP $ asymmetry can be measured. For simplicity we will call ``suppressed" ({\it sup}) this final state.
The interfering amplitudes are of the same order of magnitude, so large asymmetry effects are expected.

The direct CP asymmetry
$$
\displaystyle A_{ADS}  = \frac{\mathcal{B}(B^-\rightarrow [K^+\pi^-]_{D}K^-)-\mathcal{B}(B^+\rightarrow [K^-\pi^+]_{D}K^+)}{\mathcal{B}(B^-\rightarrow [K^+\pi^-]_{D}K^-)+\mathcal{B}(B^+\rightarrow [K^-\pi^+]_{D}K^+)}
$$ 
can be written in terms of the decay amplitudes and phases $\displaystyle A_{ADS}  =  \frac{2r_B r_D\sin{\gamma}\sin{(\delta_B+\delta_D)}}{r_D^2 + r_B^2 + 2r_Dr_B \cos{\gamma}\cos{(\delta_B+\delta_D)}}$,
where $r_B = |A(b\to u)/A(b\to c)|$, $\delta_B = arg[A(b\to u)/A(b\to c)]$ and $r_D$ and $\delta_D$ are the corresponding amplitude ratio and strong phase difference of the $D$ meson.\\
The denominator corresponds to another physical observable, the ratio between suppressed and favored ({\it fav}) events, the latter coming from the decay channel
$B^- \to D^0 K^- $ (\textit{color favored}), with $D^0 \to K^- \pi^+$ (\textit{Cabibbo favored}):
$R_{ADS}  =  r_D^2 + r_B^2 + 2r_Dr_B \cos{\gamma}\cos{(\delta_B+\delta_D)} $
$$
R_{ADS}  =    \frac{\mathcal{B}(B^-\rightarrow [K^+ \pi^-]_{D}K^-)+\mathcal{B}(B^+\rightarrow [K^-\pi^+]_{D}K^+)}{\mathcal{B}(B^-\rightarrow [K^- \pi^+]_{D}K^-)+\mathcal{B}(B^+\rightarrow [K^+\pi^-]_{D}K^+)}. 
$$

%
%
We can measure the corresponding quantities, $A_{ADS}$ and $R_{ADS}$, also for the 
$B^- \to D \pi^-$ mode, for which sizeable asymmetries may be found~\cite{ref:hfag}.

\begin{figure}[!h]
\centering
\includegraphics[width=2.9in]{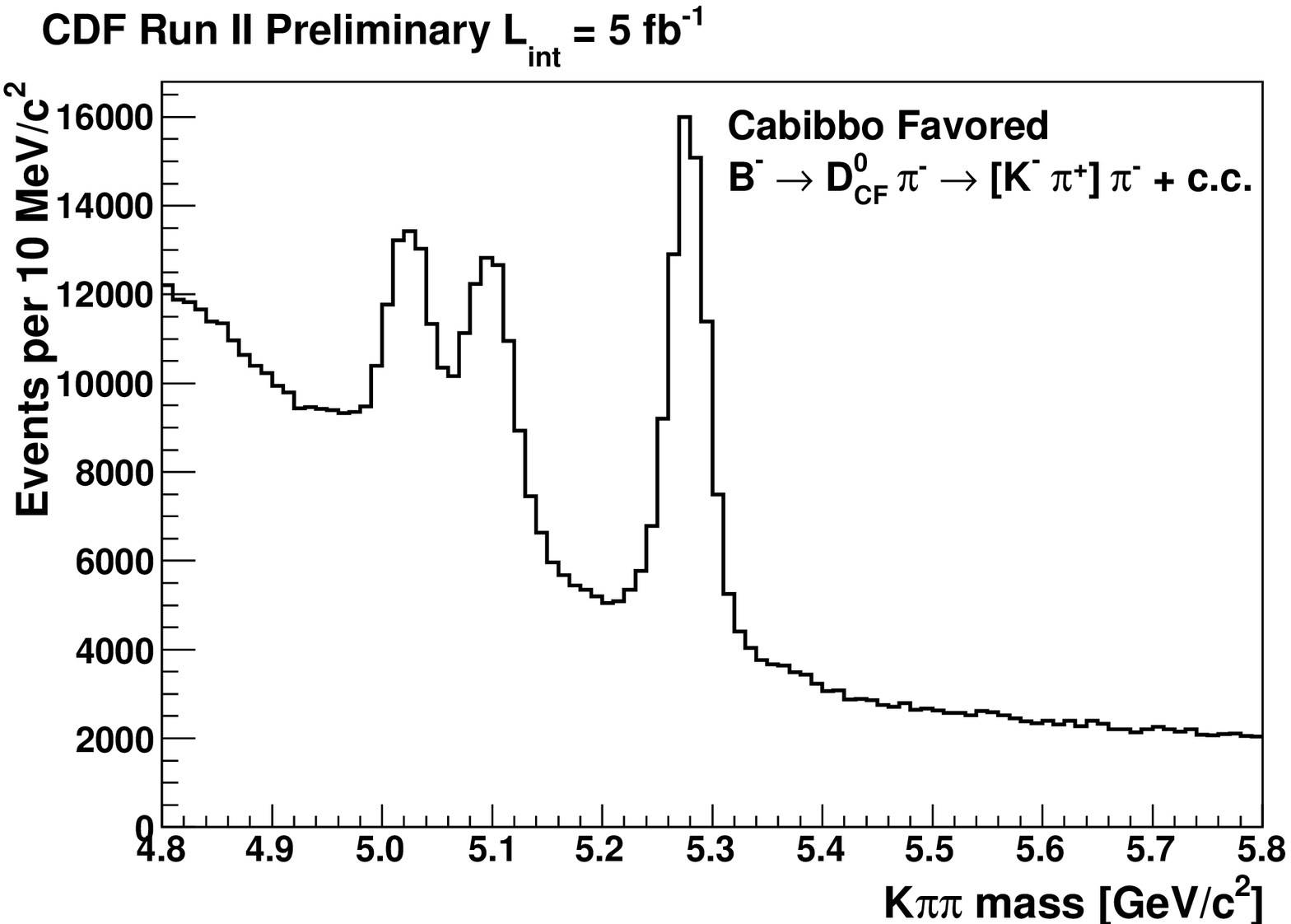} 
\includegraphics[width=2.9in]{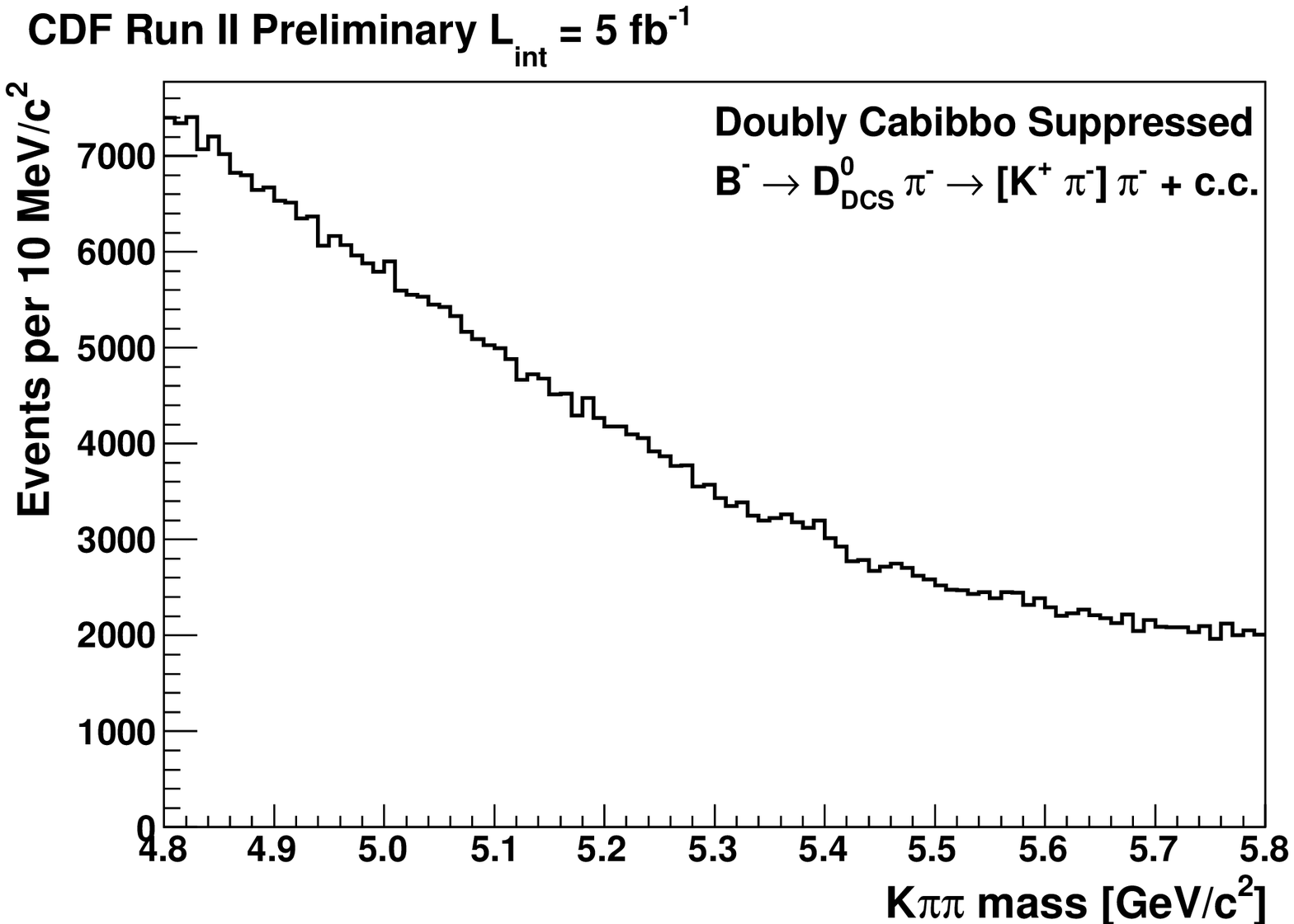}  
\caption{Invariant mass distributions of $B^- \to D h^-$ ($h$ is $\pi$ or $K$) candidates for each reconstructed
decay mode, favored on the left and suppressed on the right. } \label{fig:before_cuts}
\end{figure}

The invariant mass distributions of the favored and suppressed modes, using a sample of 5~fb$^{-1}$ of data, with a nominal pion mass assignment to the track from the $B$ meson decay, are reported in Fig.~\ref{fig:before_cuts}. 

A $B^- \to D \pi^-$ favored signal is
visible at the correct mass of about 5.279 GeV$/c^2$. 
Events from $B^- \to D K^-$ decays are expected to cluster in smaller and wider
peaks, located about 50 MeV$/c^2$ below the $B^- \to D \pi^-$ peak.
The $B^- \to D \pi^-$ and $B^- \to D K^-$ suppressed signals appear to be buried in the combinatorial background. 
Suppression of the combinatorial background is obtained through a cut optimization focused on finding a signal of the $B^- \to D_{sup} \pi^-$ mode. 
Since the $B^- \to D_{fav} \pi^-$ mode has the same topology of the suppressed one, but more statistics, we
did the optimization using signal ($S$) and background ($B$) directly from favored data, choosing a set of cuts which maximize the figure of merit $S/(1.5+\sqrt{B})$~\cite{ref:punzi}. 

\begin{figure}[!ht]
\centering
\includegraphics[width=2.9in]{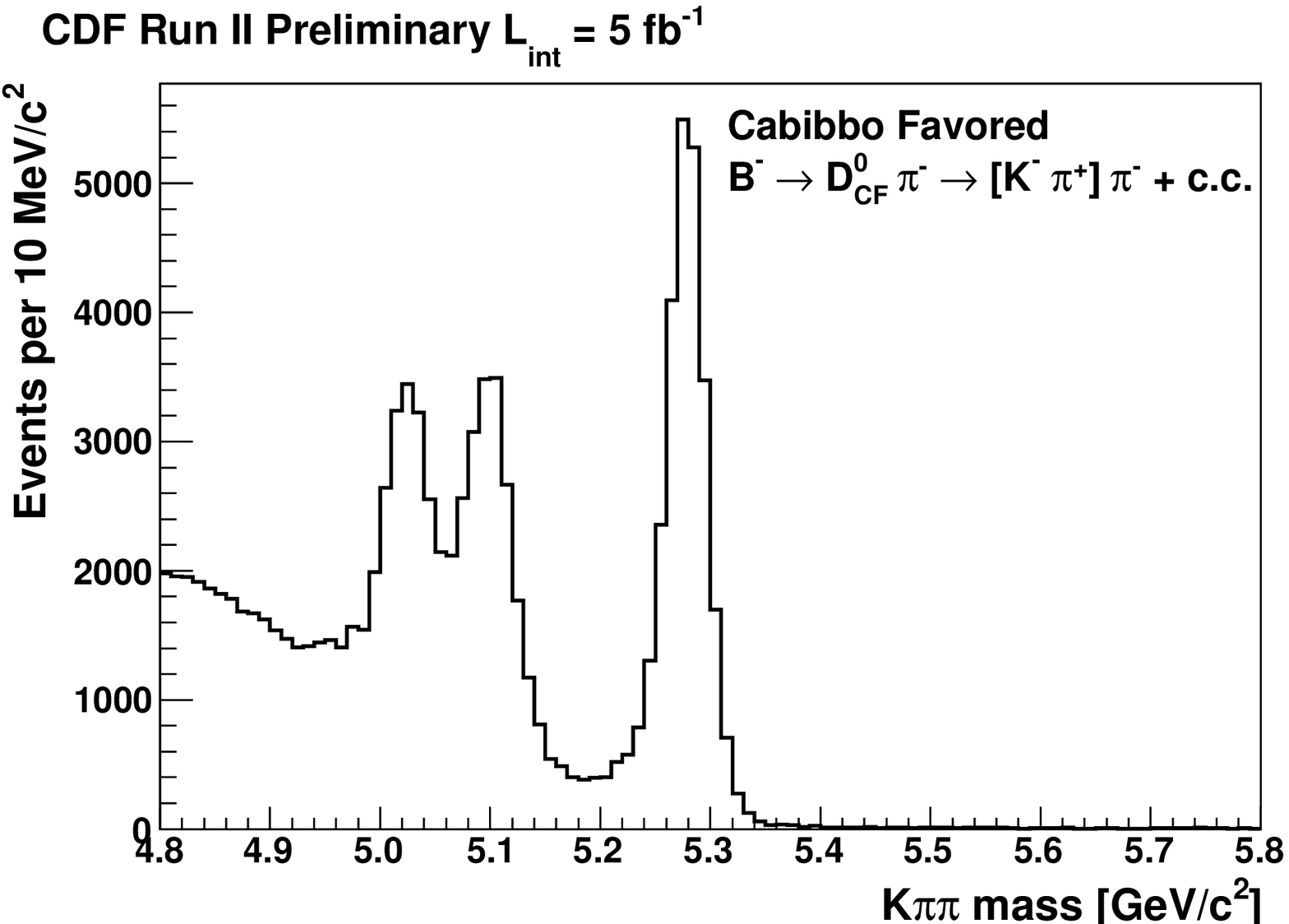}
\includegraphics[width=2.9in]{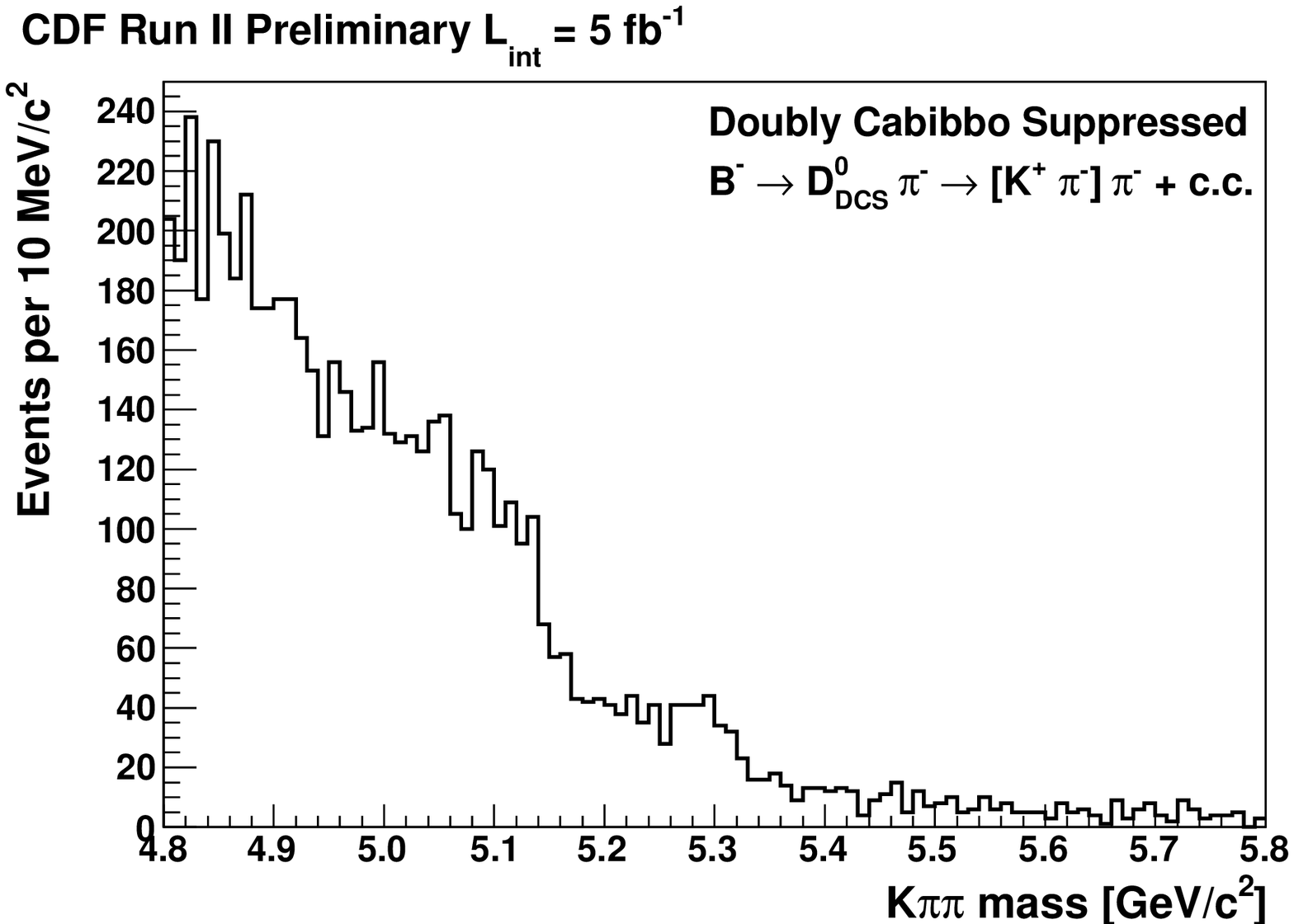} 
\caption{Invariant mass distributions of $B^- \to D h^-$ candidates for each reconstructed
decay mode, favored on the left and suppressed on the right, after the cuts optimization.} \label{fig:after_cuts}
\end{figure}

Several variables have been chosen to select signal from background~\cite{ref:pubnote}, the most important being the \textit{offline cut on the three-dimensional vertex quality} $\chi^2_{3D}$, which exploits the 3D silicon-tracking to resolve multiple vertices along the beam direction and to reject fake tracks, and the $B \ isolation$. 
Another important cut is also the \textit{decay length of the $D$ with respect to the $B$}, which allows rejection of most of the $B^- \to hhh$ backgrounds, where $h$ is either the charged $\pi$ or $K$.  
All variables and threshold values applied are described in~\cite{ref:pubnote}. 
The resulting invariant mass distributions of favored and suppressed modes are reported in Fig.~\ref{fig:after_cuts} where the combinatorial background is almost reduced in the $B^-$ mass region, allowing a signal to be seen.

An unbinned likelihood fit, exploiting mass and particle identification information is performed~\cite{ref:pubnote} 
on both favored and suppressed samples, to separate the $B^- \to D K^-$ contributions from the $B^- \to D \pi^-$ signals and the combinatorial and physics backgrounds. 
The particle identification information is provided by the specific ionization (dE$/$dx) of the CDF drift chamber which allows a $\pi/K$ separation of about $1.5\sigma$.
The dominant physics backgrounds for the suppressed mode are the inclusive $B^- \to D^0 \pi^-$, with $D^0 \to X$ (where $X$ are modes other than $K\pi$); $B^- \to D^0 K^-$, with $D^0 \to X$; $B^- \to D^{0*} \pi^-$, with $D^{0*} \to D^0 \pi^0/ \gamma$; $B^- \to K^- \pi^+ \pi^-$ and $B^0 \to D^{*-}_0 e^+ \nu_e$.

\begin{figure}[!h]
\centering
\includegraphics[width=2.95in]{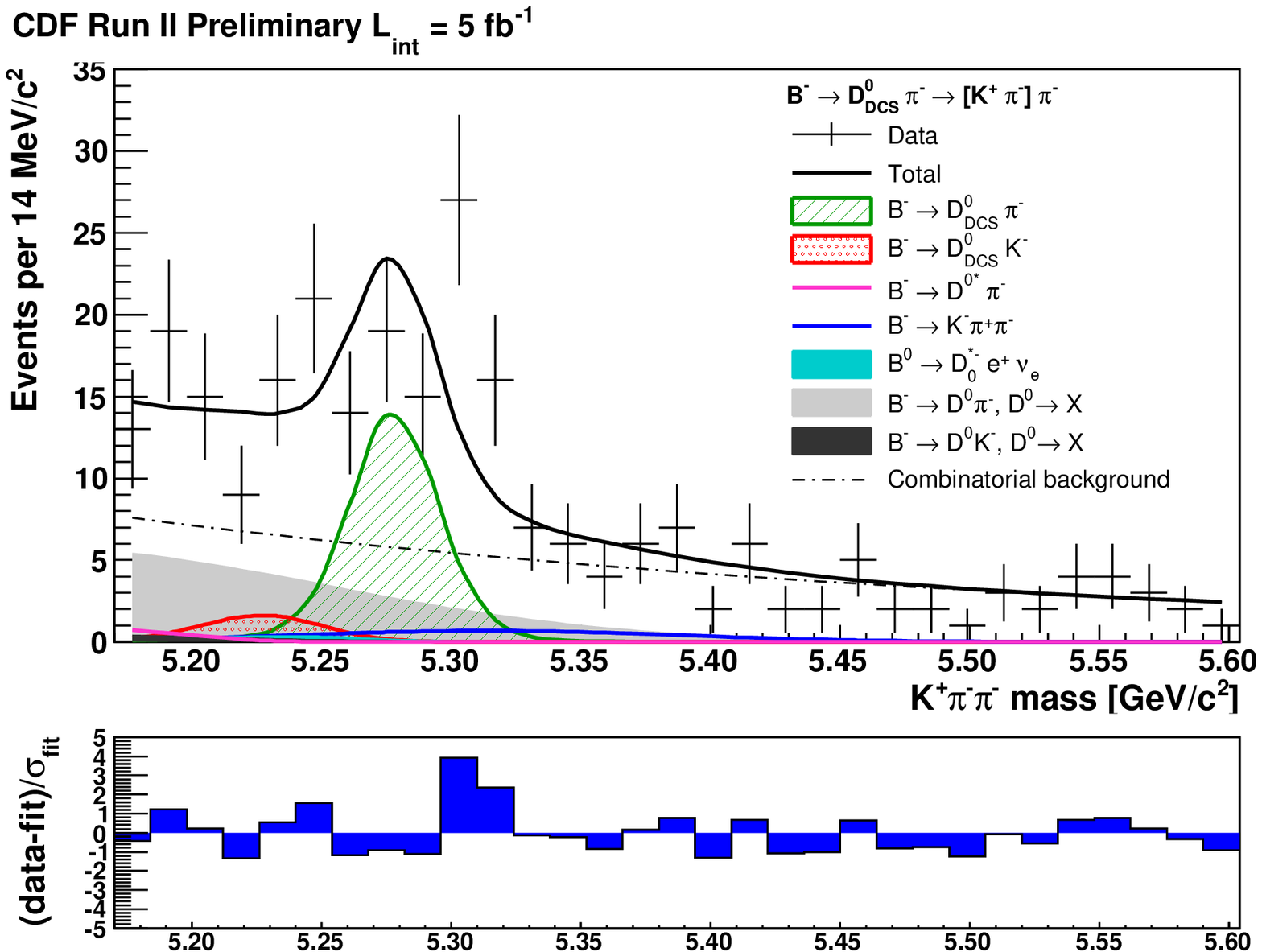} 
\includegraphics[width=2.95in]{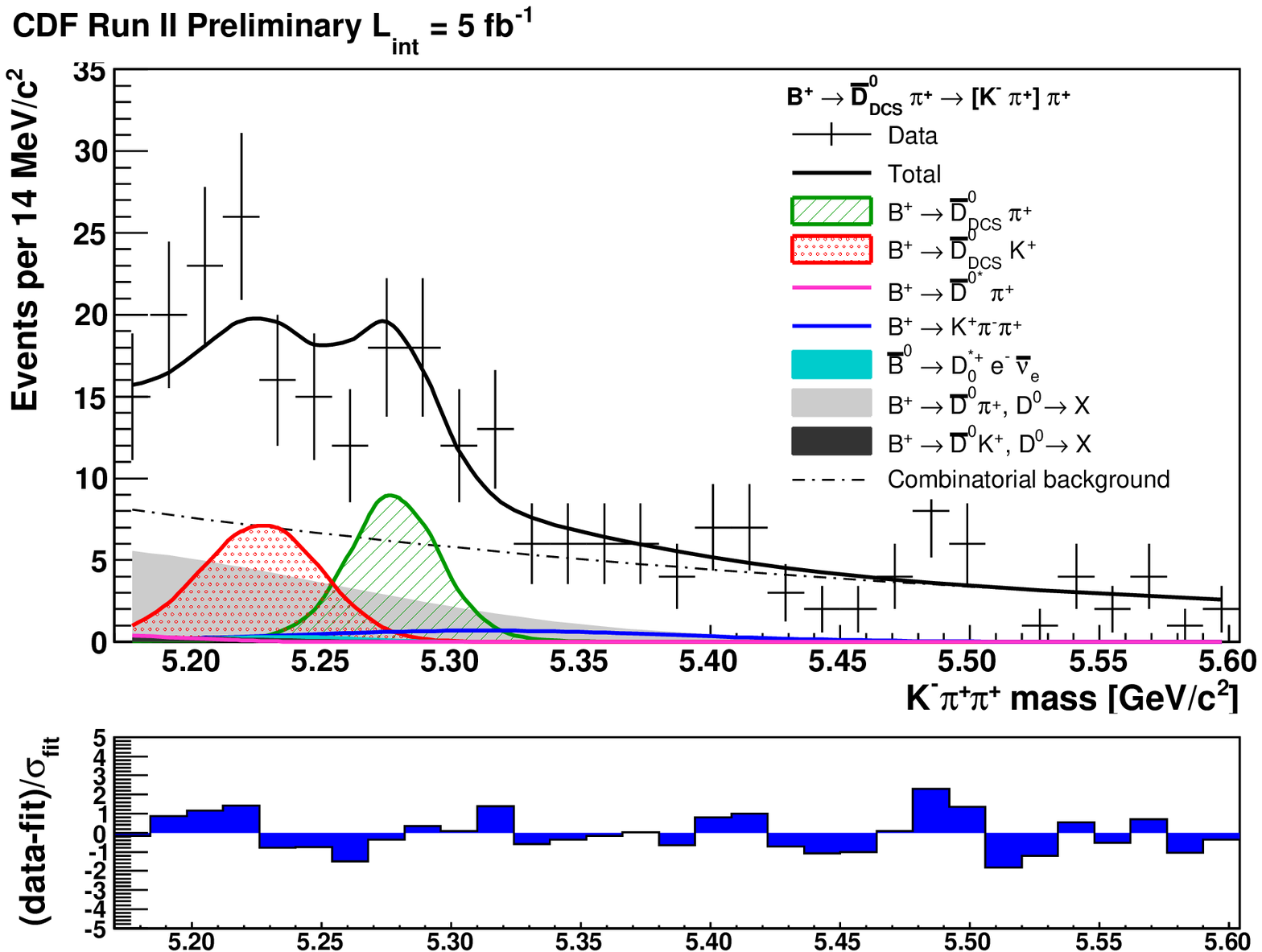} 
\caption{Invariant mass distributions of $B^- \to D_{sup} h^-$ candidates for negative (top) and positive (bottom) charges. The projections of the likelihood fit are overlaid.} \label{fig:plot_dcs}
\end{figure}

Projections of the fit in the suppressed invariant mass distributions, separated in charge, are shown in Fig.~\ref{fig:plot_dcs}.
We obtained $34 \pm 14$ $B^- \to D_{sup} K^-$ and $73 \pm 16$ $B^- \to D_{sup} \pi^-$ signal events, with a combined significance greater than 5$\sigma$.
Since $K^+$ and $K^-$ have a different probability of interaction in the detector, we evaluated the efficiency using a simulation sample and we corrected the fit results for this value.\\
The final results for the asymmetries are:
\begin{eqnarray}
A_{ADS}(K) & =  & - 0.63 \pm 0.40\mbox{(stat)} \pm 0.23\mbox{(syst)} \nonumber \\
A_{ADS}(\pi) & = &  0.22 \pm 0.18\mbox{(stat)} \pm 0.06\mbox{(syst)} \nonumber
\end{eqnarray}
and for the ratios of suppressed to favor modes: 
\begin{eqnarray}
R_{ADS}(K) & = & [22.5 \pm 8.4\mbox{(stat)} \pm 7.9\mbox{(syst)}] \times 10^{-3}  \nonumber \\
R_{ADS}(\pi) & = & [4.1 \pm 0.8\mbox{(stat)} \pm 0.4\mbox{(syst)}] \times 10^{-3}.\nonumber
\end{eqnarray}
These quantities are measured for the first time in hadron collisions and the results
are in agreement with existing measurements performed at the $\Upsilon$(4S) resonance~\cite{ref:hfag,ref:ckmfitter}.

\section{Gronau-London-Wyler method}
In the GLW method~\cite{ref:glw1,ref:glw2} the CP asymmetry of $B^- \to D_{CP\pm} K^-$ is studied, where $D$ is $D^0$ or $\overline{D}^0$ and $CP \pm$ are the $CP$ even and odd eigenstates of the $D$: $D_{CP^+} \rightarrow K^+ K^-, \pi^+ \pi^-$ and $D_{CP-} \rightarrow K^0_s \pi^0, K^0_s \phi, K^0_s \omega$.

We can define four observables: 
\begin{eqnarray}
A_{CP \pm}   &=& \frac{\mathcal{B} (B^- \to D_{CP \pm} K^-) - \mathcal{B} (B^+ \to D_{CP \pm} K^+)} {\mathcal{B} (B^- \to D_{CP \pm} K^-) + \mathcal{B} (B^+ \to D_{CP \pm} K^+)} \nonumber \\
 R_{CP \pm} &  = & 2 \cdot \frac{\mathcal{B} (B^- \to D_{CP \pm} K^-) + \mathcal{B} (B^+ \to D_{CP \pm} K^+)} {\mathcal{B} (B^- \to D_{fav} K^-) + \mathcal{B} (B^+ \to \overline{D}_{fav} K^+)} \nonumber
 \end{eqnarray}
The relations with the amplitude ratios and phases are: $A_{CP \pm}  =  2 r_B \sin{\delta_B} \sin{\gamma} / R_{CP \pm}$ and $R_{CP \pm} =  1 + r^2_B \pm   2 r_B \cos{\delta_B} \cos{\gamma} $.
Three of them are independent observables since $A_{CP+}R_{CP+} = -A_{CP-}R_{CP-}$.
Unfortunately the sensitivity to $\gamma$ is  proportional to $r_B$, so we expect to see small asymmetries.

\begin{figure}[!h]
\centering
\includegraphics[width=2.6in]{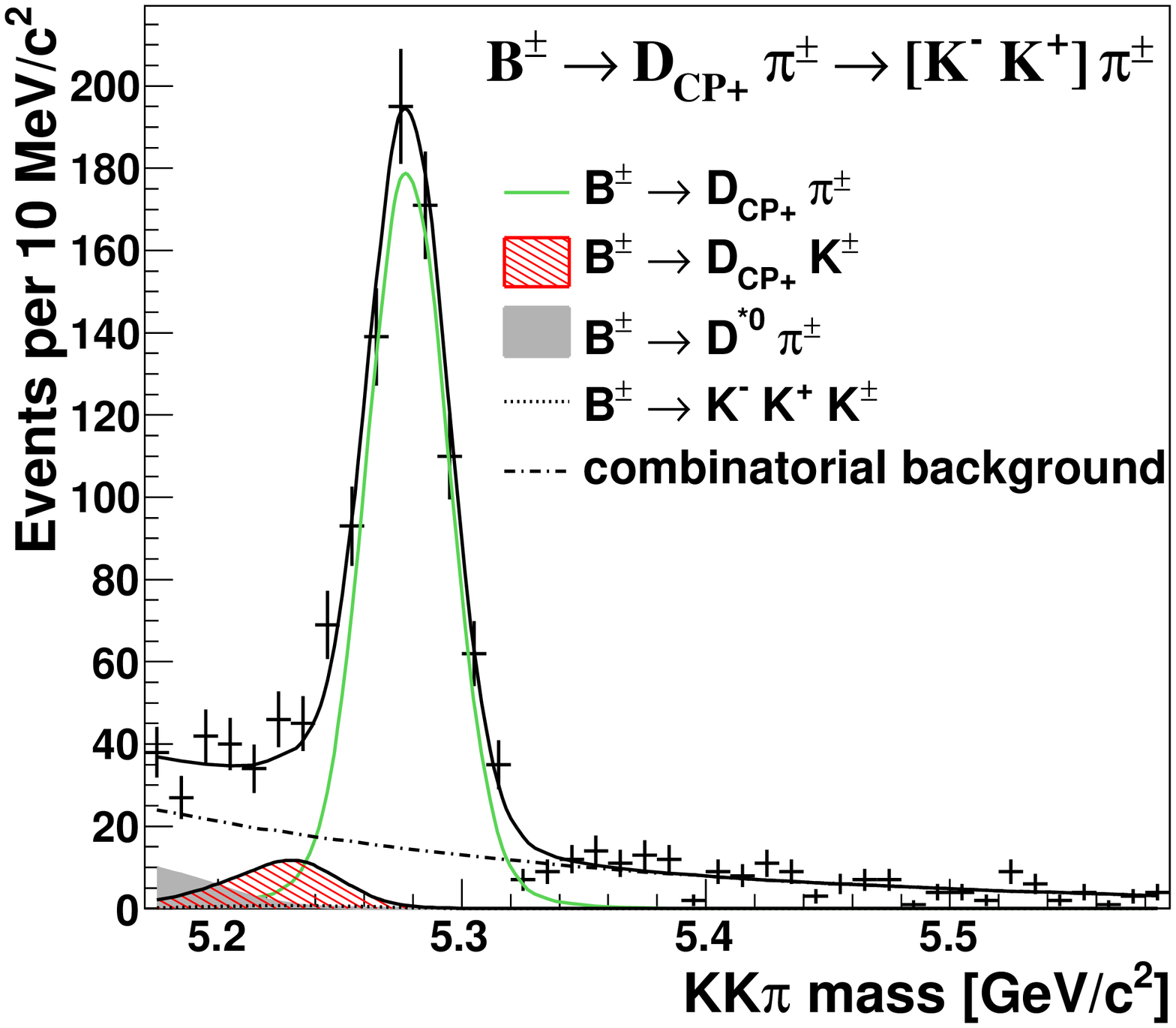}
\includegraphics[width=2.6in]{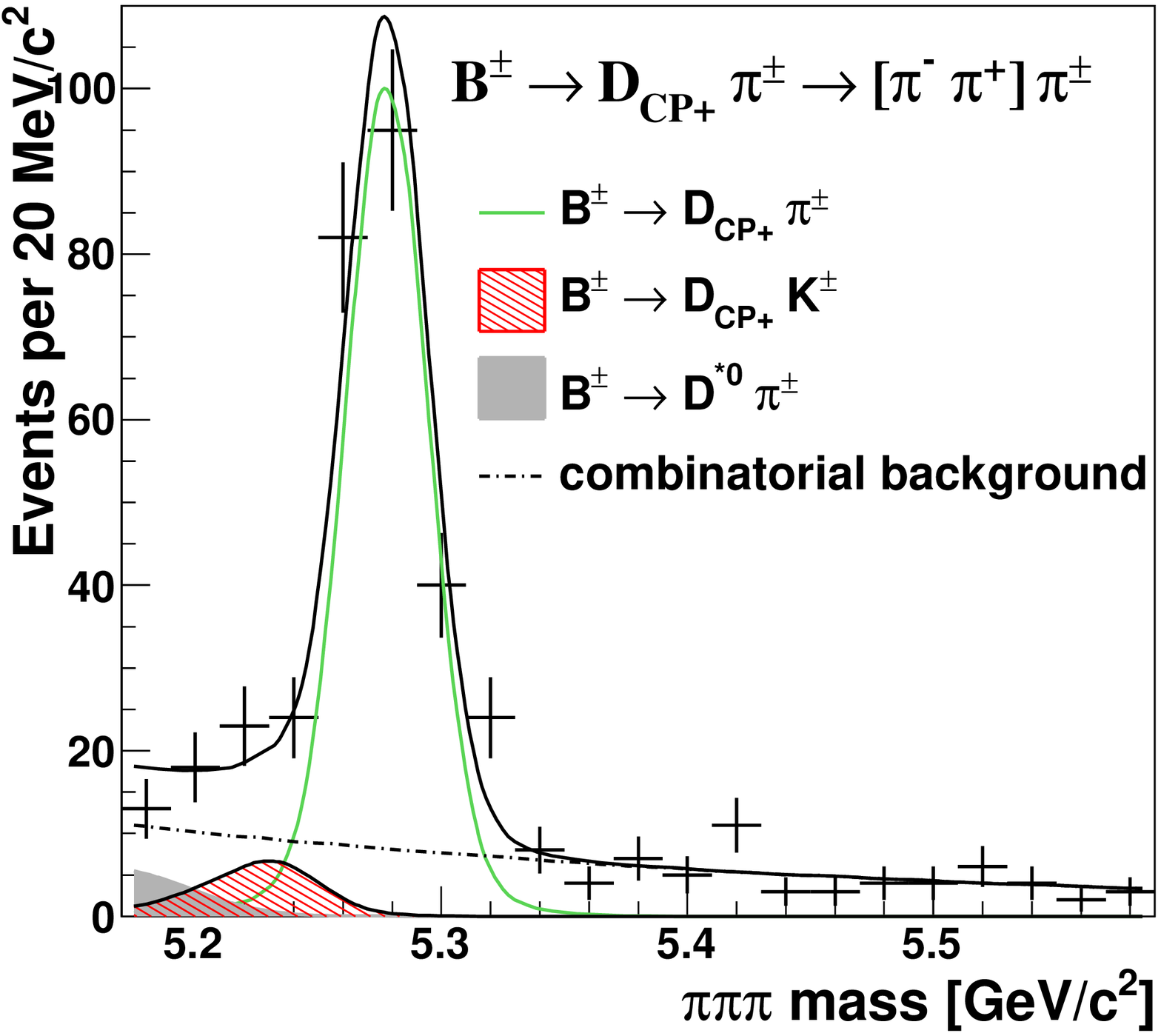} 
\caption{Invariant mass distributions of $B^- \to D_{CP} h^-$ candidates for each reconstructed decay mode, Cabibbo-suppressed $K^+ K^-$ on the left and Cabibbo-suppressed $\pi^+ \pi^-$ on the right. The projections of the likelihood fit are overlaid for each mode.} \label{fig:plots_glw}
\end{figure}

CDF performed the first measurement of branching fraction and $CP$ asymmetry of the $CP+$ modes at a hadron collider, using 1 fb$^{-1}$ of data~\cite{ref:dcp}.
The mass distributions obtained for the two modes of interest
($D \to K^+K^-$ and $\pi^+\pi^-$) are reported in Fig.~\ref{fig:plots_glw}, where a clear $B^- \to D \pi^-$ signal can be seen in each plot.

The dominant backgrounds are the combinatorial background
and the mis-reconstructed physics background such as $B^- \to D^{0*} \pi^-$ decay. In the
$D^0 \to K^+ K^-$ final state also the non-resonant $B^- \to K^-K^+K^-$ decay appears, as determined by a study
on CDF simulation~\cite{ref:notaDcp}. 
From an unbinned maximum likelihood fit, exploiting kinematic and particle
identification information, we obtained about 90 $B^- \to D_{CP +} K^-$ events and we measured the double ratio of CP-even to flavor eigenstate branching fractions and the direct CP asymmetry:
\begin{eqnarray}
R_{CP+} & = & 1.30 \pm 0.24 \mbox{(stat)} \pm  0.12 \mbox{(syst)} \nonumber \\
A_{CP+} & = & 0.39 \pm 0.17 \mbox{(stat)} \pm  0.04 \mbox{(syst)}. \nonumber
\end{eqnarray}
These results are in agreement with previous measurements from $\Upsilon$(4S) decays~\cite{ref:hfag,ref:ckmfitter}.

\section{Conclusions}
The CDF experiment is pursuing a global program to measure the $\gamma$ angle from tree-dominated processes. 
The published measurement using the GLW method and the preliminary result using the ADS method show competitive results with previous measurements performed at $B$-factories and demonstrate the feasibility of these kinds of measurements also in a hadron collider environment.
 
We expect to increase the data-set available by the end of the year 2011 and obtain interesting and more competitive results in the near future.

\end{document}